**Original Research Article**

# Impact of COVID-19 post lockdown on eating habits and lifestyle changes among university students in Bangladesh: a web based cross sectional study

Faysal Ahmed Imran[1], Mst Eshita Khatun[2]*

[1]Department of Nursing, Tairunnessa Memorial Medical College and Hospital, Gazipur, Bangladesh
[2]Department of Computer Science and Engineering, Daffodil International University, Dhaka, Bangladesh

**ABSTRACT**

**Background:** Since the confinement of the lockdown, universities transferred their teaching and learning activities in online as an all-out intention to prevent the transmission of the infection. This study aimed to determine the significant changes in food habits, physical activity, sleeping hours, shopping habits, Internet use time and mental status of the students and investigate the associations between variables.
**Methods:** The study participants were 307 Undergraduate students, between 18 and 25 years of age completed a structured questionnaire from January 3, 2022 to February 13, 2022. The questionnaire included demographic information of the students, questionnaire of dietary pattern, physical activity, sleep quality index, Shopping practice and Internet use time. Chi-square tests were used to associate the baseline information with lifestyle changes in post lockdown.
**Results:** The study reveals that 21.5% of respondents gained weight, 23.8% lost their weight and 41.7% controlled their weight. Eating of homemade food decreased after lockdown 76.5% and eating of restaurant food increased after lockdown 23.5%. A number of major meals 3-4 meals per day decreased after lockdown 61.9%. Physical exercise significantly increased after lockdown (p=0.001). Sleeping hours per day significantly decreased after lockdown (p=0.001), sleep quality was almost the same and energy level increased more in post lockdown. Respondents felt mentally tired after lockdown 60.9%. Respondents spending time on the Internet in chat rooms was 88.3%.
**Conclusions:** This study represents the significant impact on food habits, mental health, and daily routine of students after lockdown, suggesting that we should maintain a balanced diet, physical exercise to improve sleep quality and mental health.

**INTRODUCTION**

The world has experienced significant challenges with the outbreak of the novel coronavirus (COVID-19). It has been defined as a world health crisis that is profoundly responsible for the changes in social, economic and political structure. COVID-19 was first detected in December 2019 and the world health organization (WHO) declared the condition as pandemic on March 11, 2020.[1] It is detected as an international public health emergency that impacts physical, psychological, social and economical activities. As of February 5, 2022 there have been 1,844,828 confirmed cases, 1,587,374 recoveries and 28,524 deaths in Bangladesh.[2] The world researchers continuously work hard to describe the infectious agent and its variants. The epidemiological surveillance has been established to prevent and control the communicability of the disease. It is specified as a communicable disease that can spread through the droplets of an infected person's body when they cough, sneeze, speak or breathe. The major weapons used to survive against coronavirus is social distancing, self-isolation, facial mask and mobile sanitization. The Bangladesh government initially declared a lockdown to avoid gathering in public places as well as prevent the transmission of disease.

Though it is a public health emergency, People staying home for a long time are facing multiple health issues, especially Students. Educational institutions remain closed and they are the sufferers both physically and mentally due to this captivity. Their regular life and dietary patterns got significant changes due to their long period of existence at home. Unhealthy food behaviour like less vegetables and fruit consumption increases health risk as well as less intake of water makes the body dehydrated. People suffer from metabolic disorders, along with nutritional burden. More sedentary behaviour increases the chances of stroke, heart disease, obesity, diabetes, hypertension and depression. Physical inactivity becomes a major health concern for occurring life threatening disease. Prolonged sleeping patterns develop lethargy and fatigue behaviour.[3] People also suffer from mental health problems associated with anxiety, depression, psychopathy and obsession.

Consumption of food habits can influence the individual's factors such as health, social and psychological aspects. In the pandemic, people were mostly concerned about organic food and increased self-cooking rate.[4] The rapid growth has extensively influenced all age groups of students.[5,6] The university students of Al-Ahsa, Saudi Arabia significantly changed weight status (32% gained, 22% lost, 46% control) after lockdown.[7] Studies found that youths' average BMI (Body Mass Index) increased noticeably.[8] University student's group analysis observed that physical activity and sedentary working time increased in relation to different variables.[9] Sleep quality is decreased during lockdown although students spent more time in bed.[10] Changes of behaviour, dietary and performance of the students is highly influenced by worsened sleep patterns.[11,12] Worsened sleep quality is highly subjected to increased depression symptoms.[13,14] In prominent researchers found a negative impact on mentally and financial security of employees due to COVID-19 lockdown in the UK.[15] Online shopping habits increased during the lockdown period.[16,17] Due to lockdown, physical classes were restricted and online platforms were one of the media for teaching and learning. So, The COVID-19 outbreak has had a significant effect on the use of the Internet among the students.[17] This study focused on identifying the changes in life-styles including eating habits, intake of particular foods, physical activity, sleep patterns, shopping habits, Internet uses, time distribution and mental status among undergraduate students of Bangladesh after lockdown.

**METHODS**

*Study design*

A structured questionnaire was designed with the purpose of collecting data among the students of undergraduate level both qualitative and quantitative. This study is an observational and statistical approach to study qualitative factors like: dietary changes, sleeping pattern, lifestyle changes associated with lockdown. However, a questionnaire was developed from a previous study and modified in terms of Bangladeshi Students'.[18] A descriptive cross-sectional survey conducted using Google form was delivered among students over email and social media platforms like Facebook and messenger. The questionnaire is simply categorized into seven sections and each section has a particular part to carry out the study. Different socio-demographic variables like age, gender, education was included in the questionnaire. The respondents were requested to participate willingly in the study and were needed to fill responses upon receiving the link of the form. A google form was designed with the declaration of participant's consent and assurance the confidentiality of data. The respondent's data was used for only study purposes and there was no external force and identifying data in the study.

*Statistical analysis*

Variable values are represented as counts and percentages. Chi-square test was used to determine the association between variables, and t-test was used to compare the means of two groups during and after lockdown of COVID-19. A sub-analysis was also performed for weight, eating habits, physical activities, sleep quality, mental status and screen time for study between groups; by sex and by age group (18-21 and ≥25 years). Statistical analysis was performed using Microsoft Excel 2019 and SPSS version 25.0 (Chicago, IL, USA). The significance level was set $p<0.05$.

**RESULTS**

*Demographic characteristics*

A total of 307 subjects from different educational Institutions University participated in the online questionnaire. The mean age is 21.2 years; 95% confidence interval (CI), 21.1 to 21.4 (range, 18-25). The age of the participants was normally distributed; '18-21' year old, 177, in percentages 57.7%; '22-25' year old, 130, in percentages 42.3%. Among the respondents, Majority of them were female 192, 62.5% and rest of the participants were male 115, 137.5%. The educational qualifications of the respondents were categorized into 4th year, 3rd year, 2nd year and 1st year sequentially 52 (16.9%), 118 (38.4%), 65 (21.2%) and 72 (23.5%). Most of the respondents in the study stated that they have controlled their weight after post lockdown period about 128 (41.7%) and 73 (23.8%) respondents stated that they lost their weight as well as 66 (21.5%) were weight gained and rest of them were unknown.

**Table 1: Demographic data of the study respondents (n=307).**

| Variables | N | % |
|---|---|---|
| **Age (years)** | | |
| 18-21 | 177 | 57.7 |
| 22-25 | 130 | 42.3 |
| **Gender** | | |
| Male | 115 | 37.5 |
| Female | 192 | 62.5 |
| **Educational Status** | | |
| 4th Year | 52 | 16.9 |
| 3rd Year | 118 | 38.4 |
| 2nd Year | 65 | 21.2 |
| 1st Year | 72 | 23.5 |
| **Weight Status after lockdown** | | |
| Unknown | 40 | 13 |
| Loss | 73 | 23.8 |
| Control | 128 | 41.7 |
| Gained | 66 | 21.5 |

**Table 2: Eating habits during and after lockdown.**

| Variables | During lockdown N (%) | After lockdown N (%) | P value (Chi square) |
|---|---|---|---|
| **Most eaten meals during the week** | | | |
| Homemade food | 295 (96.1) | 235 (76.5) | <0.001 |
| Restaurant food | 12 (3.9) | 72 (23.5) | |
| **Number of meals per day** | | | |
| 1-2 | 80 (26.1) | 112 (36.5) | <.001 |
| 3-4 | 202 (65.8) | 190 (61.9) | |
| More than 4 | 25 (8.1) | 5 (1.6) | |
| **Tend to ignore meals** | | | |
| Yes | 135 (44.0) | 155 (50.5) | 0.10 |
| No | 172 (56.0) | 152 (49.5) | |

*Eating habits*

The relationship between changes of eating habits both during and post lockdown period where maximum respondents ate homemade food in both during and post lockdown period 295 (96.1%) and 235 (76.5%) respectively is depicted in (Table 2). However, there are significant changes of eating restaurant food where 12 (3.9%) were during lockdown that increases 72 (23.5%) in post lockdown. Among the respondents, 1-2 numbers of meals per day were significantly increased during lockdown 80 (26.1%) to post lockdown 112 (36.5%) whereas 3-4 meals and more than 4 meals per day were decreased. There were no significant changes of ignoring meals during 135 (44.0%) and 172 (56%) whereas in post lockdown period 155 (50.5%) and 152 (49.5%). Weight change and eating habits associated with different gender and age distributions in (Table 8).

The frequency of consumption of particular foods during lockdown of COVID-19 presents in (Table 3). Major food components are important to boost the immune system and fight against disease. Among the respondents, the majority of them ate 186 (60.6%) carbohydrates (bread, rice, and potato), 155 (50.5%) protein (meat, fish, chicken) and 156 (50.8%) fiber (vegetables) rich foods 3-4 times per day. About 28 (9.1%) respondents never ate sweetened foods and 50 (16%) respondents did not get in touch with soft drinks during the pandemic.

*Physical activities*

Table 4 shows that the changes of daily activities include physical exercise and household works during and after lockdown. There were significant increases of physical exercise during lockdown 135 (44%) to 172 (56%) in the post lockdown period and household activity also slightly raised from 121 (39.4%) to 135 (44%) in during and post lockdown period. In (Table 8) there is significant relation between different age groups and p=0.04.

*Shopping habits*

The obtaining results state that the tendency of online grocery shopping and stoking up on food significantly increased during lockdown 174 (56.7%). Respondents have much interest in stocking food during lockdown 148 (42.2%). Reading food level and sanitizing groceries also slightly down in the post lockdown period (Table 5).

*Sleeping patterns*

According to the study, the respondents sleeping hours per night less than 7 hours during lockdown were 106 (34.5%) whereas the rate significantly increased after lockdown to 166 (54.1%). Majority respondents slept 7-9 hours per night during lockdown that decreased 115 (37.5%) after lockdown and the rest of them slept more than 9 hours in both seasons. The rate of good Sleeping quality of the respondents during and post lockdown period almost the same respectively 214 (69.7%) and 217 (70.7%). However 61 (19.9%) respondents stated that they had excellent sleeping quality during lockdown and 60 (19.5%) had poor sleeping quality after lockdown period. The utmost part of the respondents reported that they did not face any problem in sleep during and post lockdown respectively 90 (29.3%) and 102 (33.2%). However, 49 (16.0%) respondents felt it was very difficult to sleep during lockdown and 59 (19.2%) respondents woke up frequently post lockdown. Respondents felt energized more after lockdown 66 (21.5%) than during lockdown 49 (16.0%) Where they felt laziest 87 (28.3%) in lockdown than 50 (16.3%) post lockdown.

*Mental status*

Factors that affect the mental status of the respondents include physically tired, mentally tired, resentment and sadness among the respondent. The conducted study revealed that there was no significant relationship in physical tiredness among the respondents in terms of during and post lockdown period. In (Figure 1) the significant difference in mentally tiredness of the respondents is shown. 74 (24.1%) respondents stated they felt mentally tired most of the time, which diminished 49 (16.0%) in post lockdown. Majority respondents 168 (54.7%) sometimes felt mentally tired during lockdown that raised 187 (60.9%) after lockdown. However, resentment and sadness of the respondents significantly changes in both periods.

Table 3: The frequency of intake particular foods during lockdown of COVID-19 pandemic.

| Food Items | Once/day | 3-4 times/day | More than 4 times/day | 1-4 times/week | Never |
|---|---|---|---|---|---|
| **Bread/rice/potato** | 59 (19.2) | 186 (60.6) | 35 (11.4) | 24 (7.8) | 3 (1.0) |
| **Meat/fish/chicken** | 67 (21.8) | 155 (50.5) | 30 (9.8) | 54 (17.6) | 1 (0.3) |
| **Pulse/legumes** | 69 (22.5) | 116 (37.8) | 30 (9.8) | 56 (18.2) | 36 (11.7) |
| **Milk and dairy products** | 76 (24.8) | 105 (34.20) | 42 (13.7) | 60 (19.5) | 24 (7.8) |
| **Vegetables** | 64 (20.8) | 156 (50.8) | 49 (16.0) | 33 (10.7) | 5 (1.6) |
| **Fruits** | 66 (21.5) | 112 (36.5) | 47 (15.3) | 69 (22.5) | 13 (4.2) |
| **Sweet and sweetened foods** | 60 (19.5) | 110 (35.8) | 29 (9.4) | 80 (26.1) | 28 (9.1) |
| **Soft drinks** | 58 (18.9) | 106 (34.5) | 31 (10.0) | 62 (20.2) | 50 (16.3) |
| **Energy drinks** | 45 (14.7) | 103 (33.6) | 27 (8.8) | 37 (12.10) | 95 (30.90) |

Table 4: Daily activities (physical exercise, household works) in terms of during and after lockdown.

| Variables | During lockdown N (%) | After lockdown N (%) | P value, Chi-square |
|---|---|---|---|
| **Doing physical exercise** | | | |
| Never | 78 (25.4) | 46 (15.0) | |
| Sometimes/week | 135 (44.0) | 172 (56.0) | <0.001 |
| 4-5 times/week | 50 (16.3) | 38 (12.4) | |
| Everyday | 44 (14.3) | 51 (16.6) | |
| **Doing household works** | | | |
| Never | 35 (11.4) | 52 (16.9) | |
| 1-3 times/week | 121 (39.4) | 135 (44.0) | 0.02 |
| 4-5 times/week | 50 (16.3) | 48 (15.6) | |
| Everyday | 101 (32.9) | 72 (23.5) | |

Table 5: Shopping habits in terms of during and after lockdown.

| Variables | During lockdown, N (%) | After lockdown, N (%) | P value (2 tailed test) |
|---|---|---|---|
| **Online grocery shopping** | | | |
| Yes | 174 (56.7) | 130 (42.3) | <0.001 |
| No | 133 (43.3) | 177 (57.7) | |
| **Start stocking up on foods** | | | |
| Yes | 148 (48.2) | 99 (32,2) | <0.001 |
| No | 159 (51.8) | 208 (67.8) | |
| **Reading food labels** | | | |
| Yes | 190 (61.9) | 175 (57.0) | 0.2 |
| No | 117 (38.1) | 132 (43.0) | |
| **Sanitizing/cleaning groceries** | | | |
| Yes | 282 (91.9) | 259 (84.4) | 0.005 |
| No | 25 (8.1) | 48 (15.6) | |

**Table 6: Sleeping habits in terms of during and after lockdown.**

| Variables | During lockdown, N (%) | After lockdown, N (%) | P value Chi-square |
|---|---|---|---|
| **Hours of sleep per night** | | | |
| Less than 7h | 106 (34.5) | 166 (54.1) | |
| 7-9 Hours | 154 (50.2) | 115 (37.5) | < 0.001 |
| More than 9 h | 47 (15.3) | 26 (8.5) | |
| **How would you rate your sleep quality** | | | |
| Excellent | 61 (19.9) | 30 (9.8) | |
| Good | 214 (69.7) | 217 (70.7) | < 0.001 |
| Poor | 32 (10.4) | 60 (19.5) | |
| **Did you experience any of the following** | | | |
| None | 90 (29.3) | 102 (33.2) | |
| Very difficult to sleep | 49 (16.0) | 36 (11.7) | |
| Wake up frequently | 49 (16.0) | 59 (19.2) | <0.001 |
| have difficulty to sleep | 17 (5.5) | 27 (8.8) | |
| Very early wake up and could not sleep | 28 (9.1) | 43 (14.0) | |
| Long continuous sleep | 74 (24.1) | 40 (13.0) | |
| **Describe your energy level** | | | |
| Energized | 49 (16.0) | 66 (21.5) | |
| Neutral | 171 (55.7) | 191 (62.2) | <0.001 |
| Lazy | 87 (28.3) | 50 (16.3) | |

**Table 7: Internet uses time in terms of during and after lockdown.**

| Variables | During Lockdown N (%) | After Lockdown N (%) | P value Chi-Square |
|---|---|---|---|
| **Internet uses limit** | | | |
| Once a day | 24 (7.8) | 36 (11.7) | |
| More than once a day | 33 (10.7) | 41 (13.4) | 0.23 |
| Less than once a day | 24 (7.8) | 20 (6.5) | |
| Everyday | 226 (73.6) | 210 (68.4) | |
| **Screen time for study or work (hours/day)** | | | |
| 1-2 | 103 (33.6) | 89 (29.0) | |
| 3-5 | 157 (51.1) | 136 (44.3) | 0.002 |
| >5 | 47 (15.3) | 82 (26.7) | |
| **Screen time for entertainment** | | | |
| Less than 30 min/day | 65 (21.2) | 75 (24.4) | |
| 1-2 h/day | 113 (36.8) | 137 (44.6) | 0.04 |
| 3-5 h/day | 83 (27.0) | 62 (20.2) | |
| >5 h/day | 46 (15.0) | 33 (10.7) | |

*Internet use*

The Internet uses time among the respondents during and after the lockdown is shown in (Table 7). Internet use limit has no significant relation between during and after the lockdown. Mostly students spent time on different online platforms. Among those chat rooms has been found to be very popular among students in percentage 88.3%. Otherwise, students also used social networking sites 262 (85.3%), gaming 252 (82.1%), web browsing 252 (82.1%), news 262 (85.3%) and educational platform 262 (85.3%). There is significant relationship during and after lockdown screening time for study or work (p=0.002). As well as for entertainment most of the respondents spent time 1-2 h/day during lockdown and very less number spent more than 5 hours per day. This aspect ratio significantly decreased after lockdown (p=0.04). The mostly used online platforms by participants. During lockdown period online classroom, internet TV, news, social networking and chat rooms were chosen by the participants is depicted in (Figure 2).

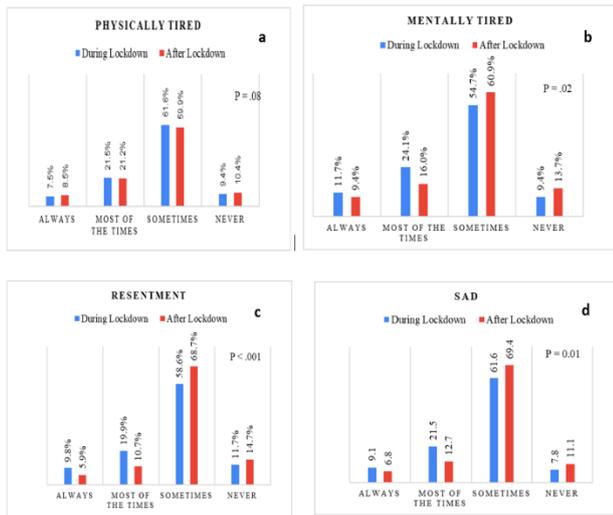

**Figure 1: Participant's mental status during and after lockdown a) physical tiredness; b) mentally tiredness; c) resentment; d) sadness.**

## DISCUSSION

The study has carried out among the students who have experienced the multiple challenges during and post lockdown period including food habit and lifestyle changes. As they are confined to practice limited actions during lockdown, adaptive behaviour and modification of lifestyle might be the possible solution to release from the difficulties resulting in post lockdown period. Prevalence of eating homemade food was higher in both during and post lockdown period whereas restaurant food was eaten more in post lockdown than during lockdown. The possible reason behind eating more restaurant food might be availability of outside food, street food as well. In the daily context of Bangladesh, people are habituated to take meals 3-4 times per day. Maximum respondents have taken 3-4 meals per day in both periods and there were no significant changes of ignoring meals. For consuming meals no association was found in the gender group but association found in age groups. Major food components like carbohydrates, protein, fibre and dairy foods are eaten 3-4 times per day. These foods were consumed to increase regular food value and boost immune power. There are also health-conscious respondents who never ate sweetened foods and did not get in touch with soft drinks during the pandemic. Healthy food habits can boost up the nutritional status and better perform against any kind of viral diseases.[19] Weight status has significantly changed ($p<0.001$) among male and female students during and after lockdown. Previous study also found that consumer's intake more calories and indicated weight gain during COVID-19.[20,21] Due to home confinement, respondents got bored and developed sedentary behaviour which may be the responsible cause for rising health burdens such as stroke, diabetes, hypertension and cardiovascular disease. The respondents were prone to physical exercise more after cancellation of lockdown. Respondents' household activities increase in the post lockdown period. Physical activity variables found association in the age group of students. In agreement with our study, in the lockdown period, a minimal number of participants reported no engagement with physical exercise. The findings of the other studies, lockdown period of COVID-19 had a notable impact on lifestyles of the humans, including decreased the physical activity and also less engagement in sports activity in general.[22] In this study, respondents had a tendency of online grocery shopping and stoking up on food during lockdown. They have much interest in stocking food during lockdown. Reading food labels read habit continuously increased among participants during lockdown. Association factor has been found in this study about sanitizing grocery items. Other studies also found online shopping is becoming familiar to the participants.[23-25]

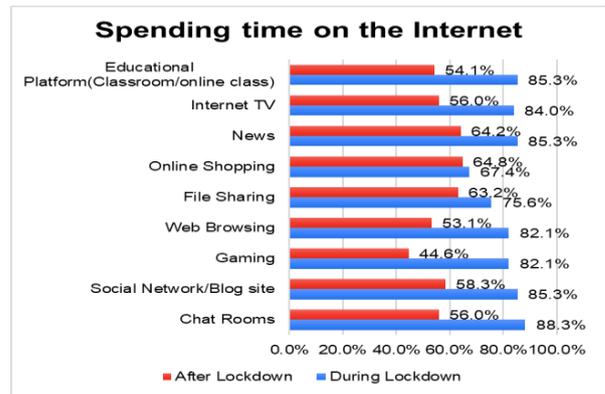

**Figure 2: Participant's most used online platforms.**

Respondents' sleeping hours per night decreased as they got activated to their regular activities after cancellation of lockdown. Most of the respondents slept 7-9 hours per night during lockdown that decreased after lockdown. Respondents' sleeping quality was almost the same in both periods. Respondents had experienced different patterns of sleeping in the pandemics. Some of them stated that they did not face any problem during and post lockdown. Whereas, 16.0% respondents felt it was very difficult to sleep during lockdown and 19.2% respondents woke up frequently post lockdown. Respondents felt energized more after lockdown than during lockdown whereas they felt laziest in lockdown than post lockdown. In this study, respondents stated that they did not feel physically tired due to staying home for a long period of time because of the restriction of movements outside. However, they reported that they felt mentally tired in both periods. The pandemic influenced the mental health of the respondents which led to anxiety, stress, frustration and depression. They felt the uncertainty about their future resulting from long term distancing from their academic activities such as classes and examination.

Imran FA et al. Int J Community Med Public Health. 2022 Jun;9(6):2449-2456**Table 8: Chi square test for the association between variables and lifestyle changes in post lockdown.**

| Variables | | Weight Change | Meals per day | Physical Exercise | Sleep Quality | Mental Status | Screen Time (study) |
|---|---|---|---|---|---|---|---|
| Gender (female, male) | DF | 3 | 2 | 3 | 2 | 3 | 2 |
| | P value | <0.001 | 0.37 | <0.001 | 0.05 | 0.02 | <0.001 |
| Age distribution (year) (18-21; 22-25) | DF | 3 | 2 | 3 | 2 | 3 | 2 |
| | P value | 0.003 | 0.03 | 0.048 | 0.31 | <0.001 | <0.001 |

About 74 (24.1%) respondents reported that they felt mentally tired most of the time, which decreased 49 (16.0%) in post lockdown. Majority respondents 168 (54.7%) stated that they sometimes felt mentally tired during lockdown that raised 187 (60.9%) in post lockdown. However, resentment and sadness of the respondents significantly changes in both periods. Authors found that sleep quality and anxiety levels have negative correlation and suggested a therapeutic strategy telepsychiatry uses.[26] This process is found helpful for providing mental health services.[27] Furthermore, the Mediterranean diet provides a protective effect to prevent the risk of cardiovascular disease and is associated with health status with decreasing mental stress and improving sleep quality.[28,29] In this study, respondents were fully dependable on using the internet in their everyday activities like online classes, assignments, meetings etc. Most of the time they had spent on different online platforms like chat rooms has been found to be very popular among students at 88.3%. Other studies also found the growing rate of Internet use in participants' daily life.[30] However, students also used social networking sites, gaming, web browsing, news and educational platforms. There is significant relationship during and after lockdown screening time for study or work (p=0.002) whereas Screen time for entertainment in computer, television, music player among the respondents spent time 1-2 hr/day after lockdown increased and respondents spent time more than 5 hours per day after lockdown were decreased. This aspect ratio significantly decreased after lockdown (p=0.04).

**CONCLUSION**

This study provides an overview to assess the eating habits, consuming foods, physical activity, sleeping habits, mental status, sleep patterns and Internet use of the undergraduate students in Bangladesh during and after COVID-19 lockdown. The results show that the attainment of taking restaurant foods increased and homemade foods decreased after lockdown. Participants consume food frequency also falling down at current time. Majority of the participants mostly took carbohydrates foods 3-4 times per day during lockdown. Participating household works slightly decreased after the shutdown period. Online grocery shopping habits, tendency to store supplementary foods and sanitizing items bought were significantly decreased in post lockdown. Energy level also boosts up after the cancellation of lockdown. Mentally tired, resentment and sadness these variables have significant association between during and after lockdown. Screening time for the study is significantly decreased in post lockdown. Weight changes. Per day taken meals, sleep quality and sleeping hours were also significantly associated with this factor. Moreover, the gender of the participants significantly associated with per day meals and sleep quality. And weight gain status found similarities in both periods due to being active in physical activities. Always staying at home encouraged association with an increased intake of food, duration of sleep, frequency of meals taken. Recommendation should be made to encourage people to increase physical activities, to reduce snack consumption, to sanitize products and try to be mentally free during future lockdown periods.

## ACKNOWLEDGEMENTS

The authors acknowledge the department of nursing, Tairunnessa memorial medical college and hospital (TMMC) who gave us continuous support and second, to all of the students who participate and help the researcher for the completion of the study.

*Funding: No funding sources*
*Conflict of interest: None declared*
*Ethical approval: The study was approved by the Institutional Ethics Committee*
**REFERENCES**

1. Cucinotta D, Vanelli M. WHO declares COVID-19 a pandemic. Acta Biomed. 2020;91(1):157-60.
2. Dynamic Dashboard for Bangladesh. Fact sheet: COVID-19. Available at: https://dghs-dashboard.com/pages/covid 19.php. Accessed on 1 April 2022.
3. Puteikis K, Mameniškytė A, Mameniškienė R. Sleep quality, mental health and learning among high school students after reopening schools during the COVID-19 pandemic: results of a cross-sectional Online Survey. Int J Environ Res Public Health. 2022;19(5):2553.
4. Najmul H, Bao Y. Impact of e-Learning crack-up perception on psychological distress among college students during COVID-19 pandemic: a mediating role of fear of academic year loss. Child Youth Serv Rev. 2020;118(2020):105355.
International Journal of Community Medicine and Public Health | June 2022 | Vol 9 | Issue 6    Page 2455